\documentclass[%
reprint,
superscriptaddress,
 amsmath,amssymb,
prapplied,
]{revtex4-2}

\usepackage{graphicx}
\usepackage{dcolumn}
\usepackage{bm}
\usepackage{hyperref}
\hypersetup{
    colorlinks = true,
    urlcolor   = black,
    citecolor  = blue,
}

\usepackage{upgreek}
\usepackage{siunitx}
\usepackage{color}

\newcommand{\mm}{\si{\,\micro\metre}}
\newcommand{\mJ}{\si{\,\micro J}}

\newcommand{\RomanNumeralCaps}[1]
\linenumbers

\newcommand{\dt}{\text{$\Delta t$ }}
\newcommand{\sX}{\text{$\sigma_{xx}$}}
\newcommand{\sZ}{\text{$\sigma_{zz}$}}
\newcommand{\sXZ}{\text{$\sigma_{xz}$}}

\begin{document}

\preprint{APS/123-QED}

\title{Bullseye focusing of cylindrical waves at a liquid-solid interface}

\author{Ulisses J. Guti\'errez-Hern\'andez}
\affiliation{Instituto de Ciencias Nucleares, Universidad Nacional Aut\'onoma de M\'exico, Apartado Postal 70-543, 04510 Cd. Mx., M\'exico.}
 
\author{Hendrik Reese} 
\affiliation{Department for Soft Matter, Institute of Physics, Otto-von-Guericke Universit\"{a}t Magdeburg, Universit\"{a}tsplatz 2, 39106 Magdeburg, Germany}

\author{Claus-Dieter Ohl}
\email{claus-dieter.ohl@ovgu.de} 
\affiliation{Department for Soft Matter, Institute of Physics, Otto-von-Guericke Universit\"{a}t Magdeburg, Universit\"{a}tsplatz 2, 39106 Magdeburg, Germany}

\author{Pedro A. Quinto-Su}
\email{pedro.quinto@nucleares.unam.mx} 
\affiliation{Instituto de Ciencias Nucleares, Universidad Nacional Aut\'onoma de M\'exico, Apartado Postal 70-543, 04510 Cd. Mx., M\'exico.}

\date{\today}

\begin{abstract}
Two converging and superimposing shock and Rayleigh waves are generated on a glass substrate by focusing laser pulses on two concentric rings in a bullseye configuration (67\mm~and 96\mm~radii). We study experimentally the threshold for substrate damage as a function of the number of repetitions and the delay (0-20\,ns). The bullseye focusing experiments are compared to a single focusing ring. Additionally, fluid-structure interaction simulations using a Volume-of-Fluid framework are utilized to estimate the stresses. The lowest number of repetitions to attain surface damage is found for constructive superposition of the Rayleigh waves, i.e., here for a delay of $10\,$ns. 
The observed damage is consistent with the simulations where the largest positive stresses ($\sim 5.6\,$GPa) are achieved for bullseye focusing with \dt = 10\,ns, followed by \dt= 20\,ns which corresponds to simultaneous shock wave focusing. In all these cases, the positive stresses are followed (a few nanoseconds later) by negative stresses that can reach $\sim-6.4\,$GPa.
\end{abstract}


\maketitle


\section{Introduction}\label{Sec: Intro}

Inertial confinement fusion is a prominent example where convergent shock waves cause extreme states of matter \cite{niu_1996, Boehly2011, Betti2016}. The shock waves may be launched by a laser generated plasma from an absorbing spherical shell \cite{Larson2004}. The group of K. Nelson~\cite{Pezeril2011a} demonstrated that a tabletop version of shock wave focusing is able to reach pressures of up to 30\,GPa near the focus point. There, a cylindrical shock wave in water at a glass-water interface was driven by laser induced vaporization of the liquid along an annular ring. After focusing, the outgoing wave creates a large reduction of pressure that is sufficient to nucleate cavitation bubbles at the focal region \cite{Veysset2018}. This strong loading of the substrate from gigapascals of positive pressure to many megapascals of negative pressure causes material fatigue.
The strength of the initial shock waves was enhanced by replacing the initially used carbon nanoparticles in the liquid that act as a linear absorber with a thin layer of gold as a plasmonic absorber \cite{Veysset2017}. As a result, not only longitudinal shock waves in the liquid but also radially converging surface waves are launched into the substrate. In absence of a liquid film, Veysset {\em et al.}~\cite{Veysset2019} obtained that the vertical displacement of the substrate at the point of focus reaches up to one micrometer within less than $5\,$ns. Simulations reveal that at a threshold of tensile stresses of around $6\,$GPa cracks are formed in their particular glass substrate~\cite{Veysset2019}. These extreme conditions can be utilized to explore the response of matter to these high pressures, e.g., the transformation of pyrolytic graphite \cite{Veysset2015} or the pressure threshold response of plastic explosives~\cite{Dresselhaus-Cooper2020}.

Damage on the fluid-solid interface may originate from flaws, which are then grown through Rayleigh waves. Zhang {\em et al.}~\cite{Zhang2019} utilized an electric discharge to excite surface waves. Here, a single point of excitation, also called a monopole source, was used. Amplification using monopole sources can be obtained through time-delay focusing. The basic idea stemmed from a Fresnel lens and was implemented with two sets of successively launched pairs of shock waves as demonstrated by Gutiérrez-Hernández {\em et al.} \cite{UG2021}.

Here we extend that idea, replacing the two pairs of point sources with two concentric ring excitations to superposition the longitudinal waves in the liquid with the surface waves on the solid to amplify stresses in the solid substrate. Therefore, we investigate the damage to the substrate as a function of the delay between their launching and compare the effect of these double excitations with that from the individual sources. 

\section{Experimental setup}
The experimental setup is derived from our previous work for transient focusing of excitation pairs \cite{UG2021}. However, this time the two laser pulses are shaped into concentric rings with a spatial light modulator (SLM, Hamamatsu, X10468-01). 

\begin{figure}
\includegraphics[width=0.44\textwidth]{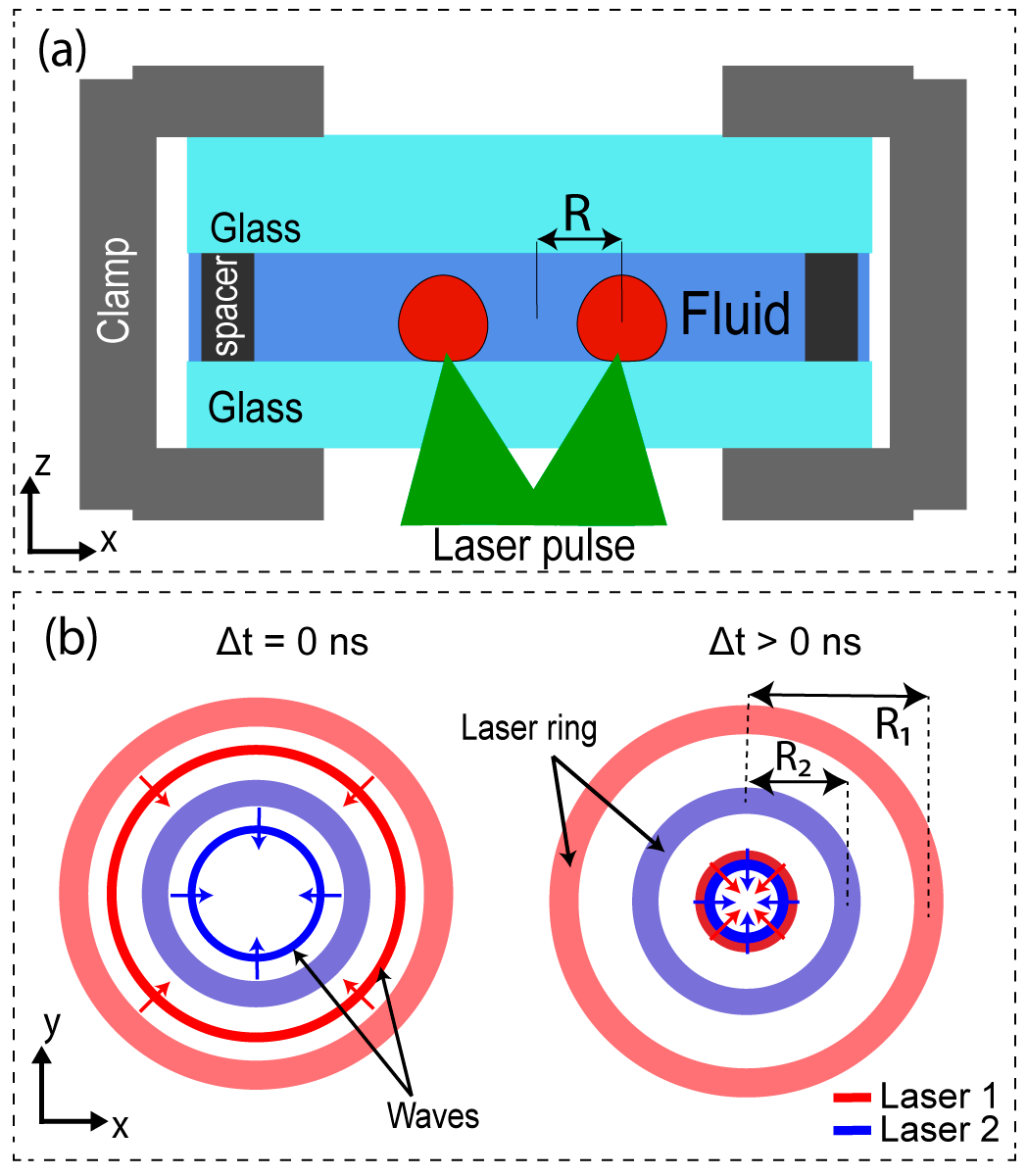}
  \caption{\label{fig: fig1} (a) cross section ($xz$ plane) of a single ring emitter. Detail of laser pulse focusing on the liquid bounded between a microscope slide and a thin cover slip. (b) Examples of (left) the two emitters launched at the same time (\dt= 0), and (right) \dt$>0$\,ns resulting in simultaneous wave focusing at the center.}
\end{figure}

Figure \ref{fig: fig1}(a) depicts a cross section ($xy$ plane) of the thin liquid sample that is bounded by two glass substrates. The top is a microscope slide (1\,mm thick) and the bottom substrate is a thin cover slip (160\mm~thick). Both substrates are made of borosilicate with elastic modulus $E=63$\,GPa, Poisson ratio $\nu=0.2$ and density $\rho=2230$\,kg/m$^3$ \cite{Fisherbrand2016a}. The height of the liquid sample is approximately 80\mm~and it is determined by spacers.

The independent beams (Nd:YAG, SOLO PIV 6\,ns, $\lambda=532$\,nm) are focused at the bottom of the container and are absorbed by the liquid (Epson printer ink, T6643 Magenta, viscosity and density are similar to water).
The absorption of the focused laser pulses lead to stress confinement \cite{Lyamshev_1981, Wu2013}. This generates cavitation bubbles with the shape of the laser pulses and shock waves are emitted because of the sudden local rise of pressure \cite{UG2021}.
The shock wave hits the solid and transfers energy to it in the form of deformation, which then becomes a bulk wave (BW) and a surface acoustic wave (SAW), or Rayleigh wave (RW), in the solid \citep{Zhang2019}. Both, the BW and RW, also induce pressure changes in the liquid.

The two laser pulses are shaped into concentric rings with an axicon phase (corresponding to a conical lens) and the radii $R_1=67$\mm~and $R_2=96$\mm. The outer ring is fired at $t=0$ and the inner ring at time $t=\dt\geq0$.

Fig. \ref{fig: fig1}(b) shows the emission from the excitations represented by the thin lines (blue for inner and red for outer), while the position of the laser beams is represented by thicker shaded lines. An example for simultaneous (\dt= 0) arrival of the laser pulses is on the left, while the right depicts a case with \dt$>$ 0 where the emitted waves are about to converge simultaneously at the center. In the drawing, the emitted waves can represent either shock waves in the liquid or Rayleigh waves in the solid.

The energy per ring is set to ($360\pm7$)\mJ. Since the two independent laser pulses have differences in their transverse profiles, with the width of the inner ring $\sim12$\mm~and that of the outer $\sim8$\mm, they have a similar fluence ($\sim7$\,J/cm$^2$).

The dynamics of the events are captured stroboscopically using a third Nd:YAG laser, (Orion, New Wave, 6\,ns, 532 nm converted to 690 nm with a dye cell) and imaged with a CCD (Sensicam QE). The red probe light is transmitted through the microscope condenser to illuminate the sample and schlieren photography is used to image the shock waves. The Rayleigh waves are not visible, likely because the deformation and change in density are so small that the created variation in brightness via a change in the index of refraction is smaller than the noise of the image.

All instruments are synchronized with a programmable delay generator (Berkeley Nucleonics 575-8C).

\section{Numerical Model}\label{Sec: Numerical Model}

The FluidStructureInteraction (FSI) package \cite{FSI2014} of the CFD software OpenFOAM \cite{OpenFOAM2016} provides the finite volume solver \textsc{fsiFoam}, which strongly couples an elastic solid and an incompressible viscous fluid.
It was modified to model a laser-generated cavitation bubble expanding close to a linear elastic boundary.
We call the new solver \textsc{CavBubbleFsiFoam}.
It has been used successfully to model the propagation of Rayleigh waves caused by a cavitation bubble in a thin liquid filled gap between two glass plates \cite{Pfeiffer2022}.

\subsection{Solid Solver}

The solid is modeled as a linear elastic material representing borosilicate glass with the Young's modulus $E=63\,$GPa, density $\rho=2230\,$kg/m$^3$, and Poisson's ratio $\nu=0.2$.
The solver \textsc{unsTotalLagrangianSolid} solves the equation of motion for a linear elastic solid:
\begin{equation}
   \frac{\partial^2\rho \boldsymbol{D}}{\partial t^2}-\nabla\cdot\Big[G\nabla \boldsymbol{D}+G(\nabla \boldsymbol{D})^\text{T}+\lambda\text{tr}(\nabla \boldsymbol{D}) 1\Big]=0\quad,
    \label{elasticEq}
\end{equation}
with the deformation field $D$, the unity matrix $1$ and the Lam\'e parameters
\begin{equation}
    \lambda=\frac{\nu}{1-2\nu}\frac{1}{1+\nu}E,~G=\frac{1}{2}\frac{1}{1+\nu}E\quad.
    \label{lame}
\end{equation}
All coordinates, at which values for deformations and stresses of the solid are given, correspond to the internal coordinates of the deformed solid.

\subsection{Fluid Solver}

Since the base fluid solver \textsc{icoFluid} is not suited to model an expanding bubble, a new fluid solver we call \textsc{CavBubbleFluid} was made to suit our purposes.
It uses code from the solver \textsc{compressibleInterFoam} which can model two compressible, viscous, non-isothermal, immiscible fluids.
It uses a phase fraction field $\alpha$ to model the fluid-fluid interface.
The fluids of interest are water and a non-condensable ideal gas describing the bubble contents.
As thermal processes exceed the scope of this study, the temperature equation is omitted and the entire domain is treated as isothermal.
Due to numerical errors, the phase fraction can fall slightly below $\alpha=1$ in the bulk of the liquid.
A correction is applied which replaces the cell values of $\alpha>0.99$ with $\alpha=1$ and of $\alpha<0.01$ with $\alpha=0$, thus removing the wrongly created gas from the liquid bulk.
The governing equations for the fluids are the law of conservation of momentum of Newtonian fluids,
\begin{equation}
   \rho\frac{D\boldsymbol{u}}{Dt}=\rho\boldsymbol{f}-\nabla p+\mu(\Updelta\boldsymbol{u}+\frac{1}{3}\nabla(\nabla\cdot\boldsymbol{u}))\quad,\label{eq:NavierStokes}
\end{equation}
and the continuity equation,
\begin{equation}
   \frac{\partial\rho}{\partial t}+\nabla\cdot(\rho\boldsymbol{u})=0\quad.\label{eq:continuity}
\end{equation}
The compressibility behaviour of the fluids is modeled via the Tait equation of state,
\begin{equation}
    p=(p_0+B)\Big(\frac{\rho}{\rho_0}\Big)^\gamma-B\quad,
    \label{tait}
\end{equation}
with the values $p_0=101325$\,Pa, $\rho_0=998.2061\,$kg/m$^3$, $\gamma=7.15$, $B=303.6$\,MPa for water and $p_0=10320$\,Pa, $\rho_0=0.12\,$kg/m$^3$, $\gamma=1.33$, $B=0$ for the gas (with $B=0$ turning the equation of state into the ideal gas equation).
Since the fluid-structure interface moves, the fluid mesh must be deformed to conform to the boundary conditions.
The dynamic mesh motion is done by solving a Laplace equation for the mesh deformation field, with the condition that the mesh boundary must coincide with the specified rigid boundaries and the deformed fluid-structure interface.

\subsection{Fluid-Structure Interaction}
The fluid exerts surface forces $\boldsymbol{F} S_i=\upsigma\cdot\boldsymbol{n_i}$ onto the fluid-structure interface via the stress tensor $\upsigma=-p1+\upmu\big[\nabla\circ\boldsymbol{u}+(\nabla\circ\boldsymbol{u})^\text{T}\big]$.
In turn, the deformation $\boldsymbol{D}$ and velocity $\dot{\boldsymbol{D}}$ of the solid boundary get imposed onto the fluid at the fluid-structure interface.
The FSI solver uses strong coupling, meaning that in each time step, the fluid solver and solid solver get solved alternatingly, until a maximum number of iterations (100) is reached or the residual between the interfaces of the fluid domain and the solid domain falls below a tolerance value. 

We analyse the horizontal stress \sX~(equal to the radial stress due to axisymmetry), the axial stress \sZ~and the shear stress \sXZ. 

\subsection{Geometry and initial conditions}

The simulated geometry describes a thin sheet of water (80\mm~height) between a rigid boundary and an elastic glass sheet (160\mm~height).
As in the experiment, a gaseous domain is put on the opposite side of the glass plate (80\mm~height), allowing it to move away from or towards the water filled domain.
Because of the rotationally symmetric nature of the experimental setup, the simulated geometry can be chosen to be axisymmetric.
It thus becomes an effectively two-dimensional problem, and only a thin wedge of a cylinder of radius 1\,mm has to be modeled.
The geometry is divided up in the horizontal and axial directions into a square mesh of computational cells with a width of 1\mm.\\
The ring bubbles are initiated with a major radius of 67\mm~and a minor radius of 12\mm~for the inner ring, and with a major radius of 96\mm~and a minor radius of 8\mm~for the outer ring.
Their centre is chosen to be at $x=$ 0\mm~and the height to be equal to the minor radius, so that the lower end of each ring touches the glass plate.
To smoothen the surface of the bubble seed it is smeared via $\alpha'-4\cdot10^{\text{-}11}\Updelta\alpha'=\alpha$.
The initial pressure is $1.6\,$GPa, which is consistent with previous works \cite{Veysset2017, UG2021}.\\
\textit{SymmetryPlane} boundary conditions are applied to the front and back of the wedge, and to the inner side representing the axis of symmetry.
The outer boundary of the fluid domain acts as an outlet and is imposed with \textit{waveTransmissive} boundary conditions.
The fluid domains are bound by either rigid or deformable solid boundaries at the top and bottom, which are treated with a \textit{zeroGradient} boundary condition for the pressure and a \textit{noSlip} boundary condition for the velocity, which means that the relative motion between the fluid at the boundary and the boundary itself it kept at 0.
For the mesh deformation, a \textit{slip} boundary condition is applied to the perimeter of the wedge geometry, whereas the deformation of the fluid-structure interface is calculated within the solver.

\section{Experimental results}
In this section we present the experimental results including the measured dynamics of the shock waves compared with the simulations. We also show photographs of the substrate damage and the statistics for repeated experiments.

\subsection{Shock wave dynamics}
\begin{figure*}\includegraphics[width=\textwidth]{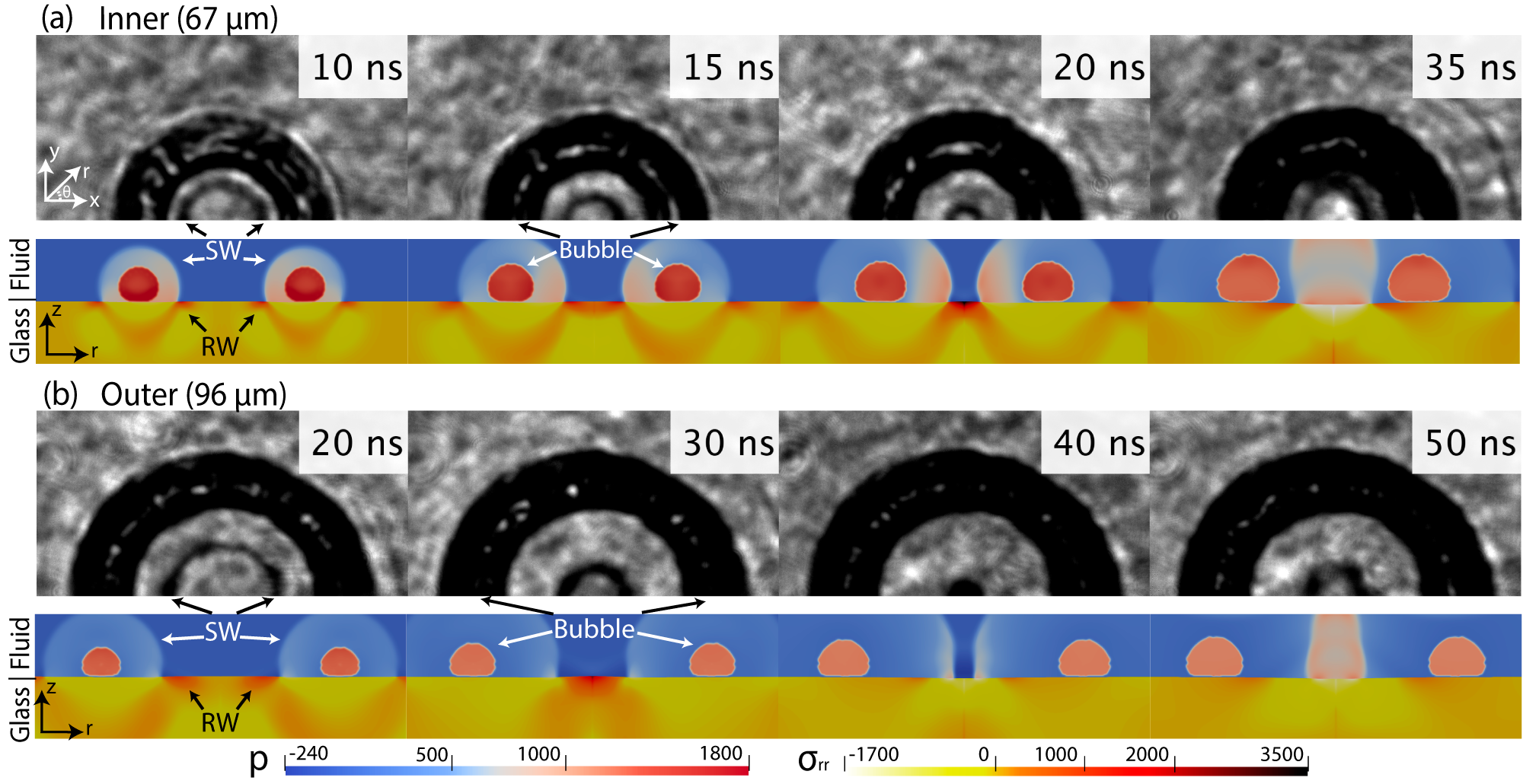}
  \caption{\label{fig: fig2} Shock wave dynamics of the two rings. The first row shows the strobe photographs ($xy$ plane), while the second shows the frames extracted from the simulation ($xz$ plane). The size of each strobe photograph is 150\mm~$\times$ 300\mm~($xy$ plane). In the results of the simulation, above is the fluid and below the glass. The size of each simulation frame is 100\mm~$\times$ 300\mm~($xz$ plane). Color bars represent pressure (left) and \sX~(right), both in MPa. (a) Inner ring. (b) Outer ring. }
\end{figure*}

Selected results for the inner and outer ring laser excitation are shown in Fig. \ref{fig: fig2}(a) and (b), respectively.
The respective top row of Fig. \ref{fig: fig2}(a) and (b) shows strobe photographs ($xy$ plane) of the events caused by the generation of the laser ring excitation at the sample, while the bottom row shows the simulation results ($xz$ plane) for the same time. Notice that the strobe photographs are cropped ($xy$ plane, $y>0$) to have a better comparison with the simulation.

The thick dark ring in the strobe photographs is the expanding cavitation bubble, while the thinner ring that propagates towards the center is the converging shock wave.
The bottom rows of Fig. \ref{fig: fig2}(a) and (b) show the corresponding simulated events for the inner (Fig. \ref{fig: fig2}a) and outer ring (Fig. \ref{fig: fig2}b). The simulations show a cross-section of the plane $xz$. The glass boundary is at $z=0$\mm, at half height of the frame. The upper half of the simulated frames contains the liquid, where we observe the bubble ring and the shock wave. In the solid, we can observe the surface waves propagating at a higher speed than the shock waves in the fluid. 

The Rayleigh wave is marked by arrows in the first frame of Fig. \ref{fig: fig2}.
The position of the Rayleigh waves was extracted by finding the maximum value of the horizontal stress, \sX, until the moment where the waves converge at the center, where it is no longer possible to track the wave once it passes through itself at the center due to interaction with the stresses induced by the incoming shock waves.
On the other hand, tracking the shock wave after it pass through itself at the center can be done as in the liquid the induced pressures due to the surface waves is very small.

\begin{figure}\includegraphics[width=0.49\textwidth]{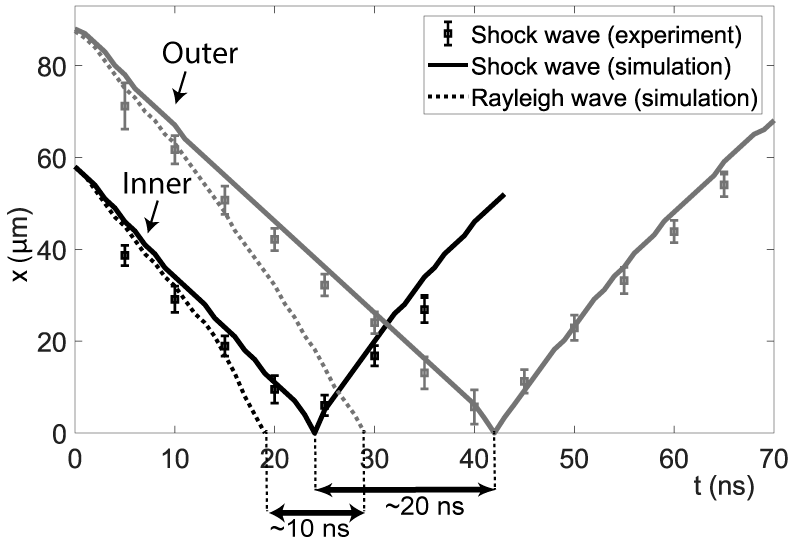}
  \caption{\label{fig: fig3} Shock and Rayleigh wave dynamics. The symbols show the measurements of the shock wave. The continuous lines represent the position of the shock waves extracted from the simulations ($p_\text{max}$ in the fluid). The dotted line represents the position of the Rayleigh wave from the simulation (\sX$_\text{,max}$ in the glass). Gray for the outer, black for the inner.}
\end{figure}

The position of the shock and Rayleigh waves over time is plotted in Fig. \ref{fig: fig3}. The waves emitted from the inner ring are drawn with black symbols and lines, and the outer ones in gray color. 
The continuous lines show the positions of the simulated shock waves over time, while the dashed lines represent the dynamics of the Rayleigh waves on the surface of the substrate extracted from the simulation. 

In the simulation, the inner Rayleigh and shock waves arrive at the origin at 19 and 24\,ns and the outer ones at 29 and 42\,ns, respectively. Assuming no interactions between inner and outer excitations, the difference in time between the Rayleigh waves is 10\,ns, while for the shock waves the time difference is 18\,ns.
 
Figure \ref{fig: fig3} shows that there is a reasonable agreement between the numerical simulations and the experimental results for the positions of the shock waves (speed $\sim$\,1800\,m/s), while the initial speeds of the Rayleigh waves is $\sim$\,3000\,m/s for the inner and $\sim$\,3100 m/s for the outer ring, which agree with the expected value of $v_\text{RW}$ = $\sqrt{E/(2\rho(1-\nu)}(0.862+1.14\nu)/(1+\nu)=3116\,$m/s \cite{Achenbach}.

Furthermore, in the simulation we also identified other surface waves like the bulk wave (BW) that propagates at $\sim 5350\,$m/s which is in reasonable agreement with the expected speed of 5602\,m/s. In this work we do not consider the induced stresses by the BW because these are a factor $\sim 30$ smaller than those of the Rayleigh waves. Also, the pressures that are induced in the liquid by the waves in the solid are much smaller than those induced by the convergence of the shock waves. The bulk wave (BW) generates negative stresses (compression) in the solid, inducing positive pressures in the liquid that are between 45-49 times smaller than the pressures reached during shock wave convergence. The Rayleigh wave propagates by generating positive stresses (tension) inducing negative pressures in the liquid that are $\sim 17$ times smaller than those induced by the shock waves.

\subsection{Substrate damage}
The experiments are done for the cases of individual rings (inner/outer) and the bullseye configuration with a time delay between both rings \dt= 0, 10 and 20$\,$ns where the outer ring is fired first at $t = 0$ and the inner ring at time $t=\Delta t$.
For each case, the experiments are performed repeatedly at a rate of 0.5\,Hz in pairs where the first shot fires the excitations and then one second later a photograph is acquired in bright-field to analyze the damage after the bubbles have disappeared. 

\begin{figure}
\includegraphics[width=0.46\textwidth]{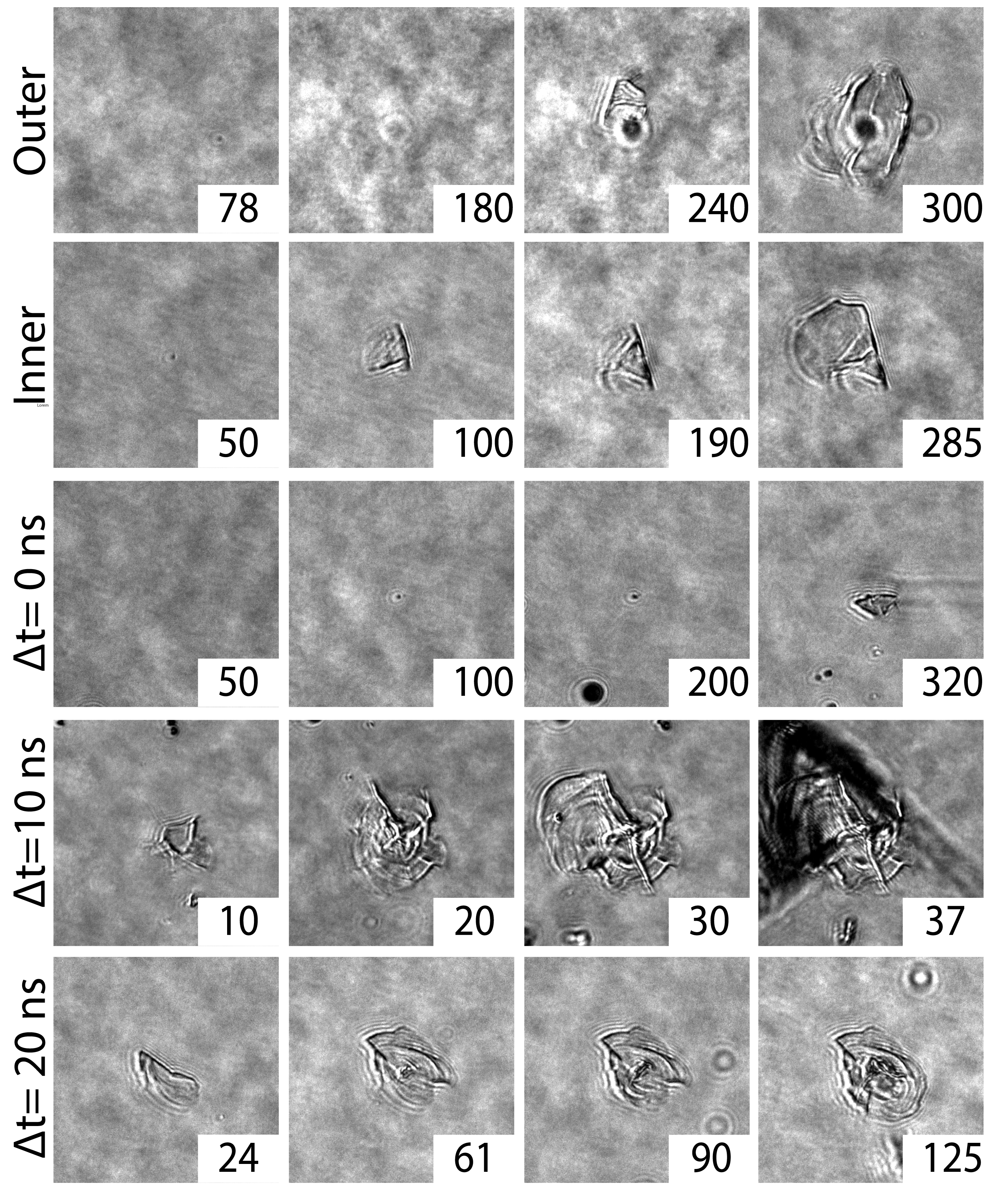}
  \caption{\label{fig: fig4}Damage as a function of the number of repetitions. The first row is with only the outer ring. Second row with only the inner ring. Third row: simultaneous firing of both rings. Fourth row: Time delay of 10\,ns (superposition of inner and outer Rayleigh waves at the center). Last row: Time delay of 20\,ns (superposition of inner and outer shock waves at the center). Each individual frame has a width of 150\mm.}
\end{figure}

Typical examples of the progression of damage at one position for repeated firing of the laser ring excitations are shown in Fig. \ref{fig: fig4}. 
The different rows represent the five cases, and show the respective developments of the damage in terms of the number of repetitions (labeled in the bottom right corner of each image).

We observe that for the outer ring (first row), damage is visible after $\sim 240$ repetitions, while in the case of the inner ring (second row) the number of repetitions to observe damage is lower at about 50. 

It is worth noting that in the case of the outer ring, a blurred circle starts to emerge in the second frame (after $\sim 180$ repetitions), which was identified as damage to the upper glass substrate.
This indicates that high pressures are achieved that can induce high enough stresses to damage the upper solid boundary. We only observe that kind of damage (upper substrate) for the individual outer ring.

Surprisingly, in the case of the two rings appearing simultaneously (third row), the damage is smaller than that produced by the individual inner ring for a similar number of repetitions.
The case of \dt= 10\,ns (fourth row) is the one that shows the strongest damage to the substrate with a smaller number of repetitions.
This is followed by the delay of \dt= 20\,ns, in the fifth row.

\begin{figure}
\includegraphics[width=0.49\textwidth]{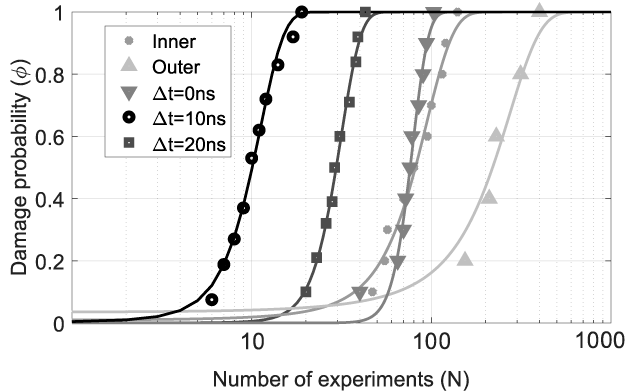}
  \caption{\label{fig: fig5} Probability of glass damage as a function of the number of repetitions for the five different conditions: outer ring, inner ring, and two rings with time delays of 0, 10 and 20\,ns. The continuous lines are fits to the cumulative distribution functions of the standard normal distribution.}
\end{figure}

To measure the probability of damage as a function of the number of repeated excitations we start the experiment in an undamaged region of the substrate, then the excitations are fired $N$ times, where $N$ varies between $20$ (for \dt= 10\, ns) and $400$ (individual outer ring) depending on the configuration (individual or bullseye) which changes the rate at which the substrate is damaged. 
Then, the substrate is displaced to expose an undamaged area to another $N$ excitations. The process is repeated $n$ times so that $n\cdot N = 2000$, with $n$ between 5 and 100 (smallest for the outer ring, largest for \dt$=10$\,ns). The probability for observing damage is $\phi (N)=\sum _{i=1}^{n} p_i(N)/n$, with $p_i(N)= 1$ when damage is observed (at the $N$ shot) and $p_i (N)= 0$ with no visible damage at the $N$ excitation. Notice that the experiments are independent for each condition, which is why we have to adjust the number of repetitions $N$ at a single position and the number of sampled positions $n$ in order to have enough points for $\phi(N)$.

Figure \ref{fig: fig5} shows the probability of observing damage ($\phi$) in the glass as a function of repetitions ($N$) for the five different cases (individual outer and inner and both with \dt= 0, 10, and 20\,ns). The symbols represent the experimental data and the continuous lines are fitted cumulative distribution functions of the standard normal distribution for each case,
\begin{equation}
    \phi_\text{fit}=\frac{1}{2}\Big(1+\text{erf}\Big(\frac{N-a}{b\sqrt{2}}\Big)\Big)\quad,
\end{equation}
where $a$ is the mean, $b$ the standard deviation and erf is the error function. Table \ref{tab:table1} contains $a$ and $b$ for all cases. We find the smallest mean value (10.16) for the case \dt= 10\,ns, while the largest is for the individual outer ring (219.27). The values are similar for the individual inner ring (81.54) and for \dt= 0 (76.23). However, the observed damage after some repetitions, subsequent to the first damage, is larger for the inner ring.

\begin{table}
\begin{ruledtabular}
\begin{tabular}{ccccccc}
 & Inner &Outer&\dt= 0\,ns & \dt= 10\,ns&\dt= 20\,ns&\\ \hline
 a & 81.54 & 219.27  & 76.23 &  10.16  & 29.55 \\
 b & 34.51 & 120.88   & 14.46 &   3.24   & 7.72  \\
\end{tabular}
\end{ruledtabular}
\caption{\label{tab:table1} Fitting parameters for a Gaussian distribution of the number of repetitions at which damage occurs, for the five cases studied. Mean value ($a$) and standard deviation ($b$).}
\end{table}

\section{Numerical Results and discussion}\label{Sec: discussion}
In this section we show the numerical results including the deformation of the solid, the calculated stresses for the individual rings and for the bullseye configuration and discuss the results considering the experimental observations.

\subsection{Details of the solid deformation}
\begin{figure}
\includegraphics[width=0.49\textwidth]{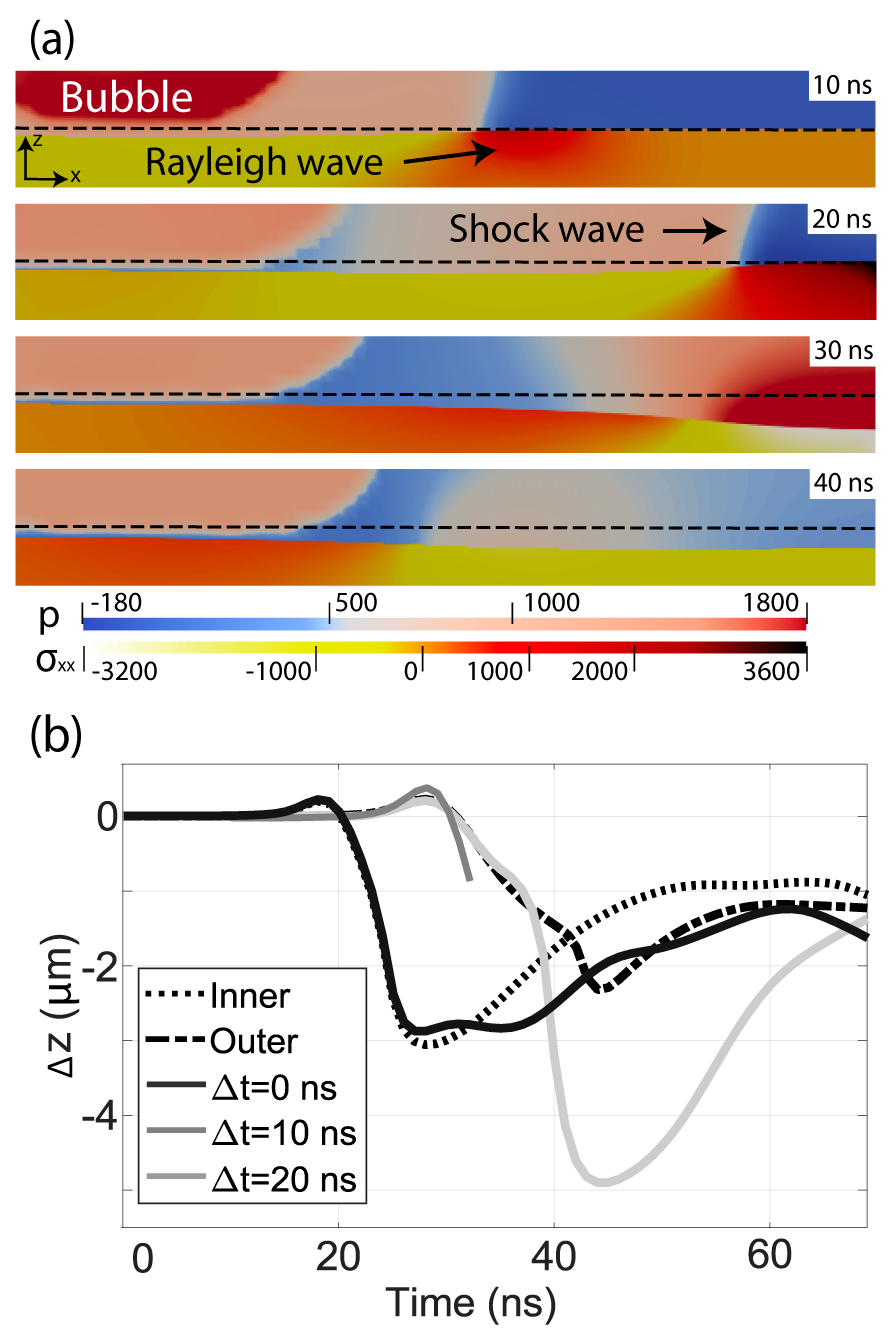}
  \caption{\label{fig: fig6}Solid deformation. (a) Frames extracted from the simulation for the inner ring. The dashed line represents the original position of the solid boundary ($z=0$). The height of the frames is 10\mm~and the width is 75\mm. The horizontal coordinate goes from $x=0$ to $x=75$\mm. The color bars represent pressure (upper) and \sX~(lower) both in MPa. (b) Deformation in the axial dimension ($\Delta z$ of the solid at the center ($z=0$, $x=0$) for all focusing conditions.}
\end{figure}

Snapshots from the simulation of the individual inner ring are shown in Fig. \ref{fig: fig6}(a), where pressures and stresses are plotted in the $xz$ plane for the inner ring. 

Figure \ref{fig: fig6}(b) shows the deformation in the axial dimension $\Delta z=z-z_0$ of the solid surface at the center ($x=0$, $z=z_0=0$) for all focusing configurations.
We observe that a lesser deformation is reached for the outer ring, with a minimum value of about -2.3\mm. The inner ring results in $\Delta z$ reaching $\sim-3$\mm~and the result for $\Delta t=0$ is similar up to 27\,ns but $\Delta z$ does not reach $-3$\mm. The largest $\Delta z$ is close to -5\mm~for \dt= 20\,ns. The case for \dt= 10\,ns does not converge past 33\,ns. However, we observe that it reaches the largest positive $\Delta z$. It is also worth noting that in all cases due to the excess pressure in the liquid, the glass plate keeps being pushed downwards over the course of the simulation.
Unsurprisingly, this only occurs in places the shock wave has already passed over.
For example, in the case for the inner ring (black dotted line in Fig. \ref{fig: fig6}b) $\Delta z=-3$\mm~at 29\,ns due to the focusing of the shock wave, but later the substrate does not return to its original position ($\Delta z =0$) but reaches $\Delta z \sim-1$\mm. At later times $t\sim$\,85\,ns (not shown in the plot) there is another negative peak due to the rebound of the shock wave. After that, the glass is still pushed downwards in its entirety due to the enduring overpressure. At $t=$150\,ns the displacement has reached -3.39\mm. 

\subsection{Individual rings}
\begin{figure}
\includegraphics[width=0.45\textwidth]{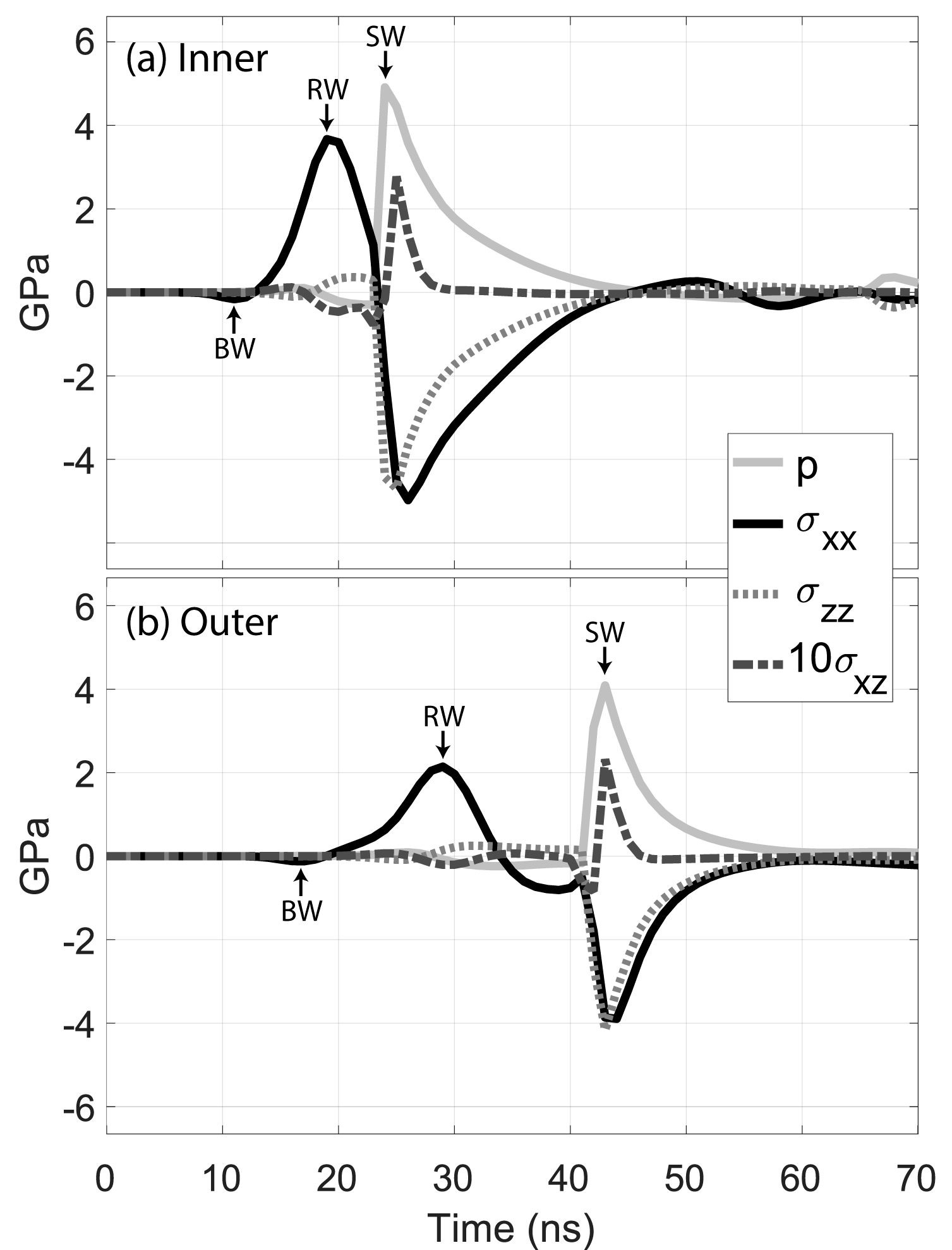}
  \caption{\label{fig: fig7}Calculated pressures and stresses at the center of the rings ($x=0$) at the solid/liquid interface ($z=0$). The arrows indicate the peaks associated to the arrival of the bulk waves. (a) Inner ring. (b) Outer ring.}
\end{figure}

The calculated pressures and stresses at the center of the rings ($x=0$) on the fluid-structure interface in the case of individual rings are plotted in Fig. \ref{fig: fig7}.
In the plots, the gray continuous line shows the pressure in the liquid at the fluid-structure interface at the origin ($x=z=0$). The peak of the pressure curve indicates the shock wave converging at the axis of symmetry.
However, prior to shock wave convergence, there is a slight increase in the pressure due to the previous passage of the bulk wave and a subsequent slight decrease to negative values due to the Rayleigh wave.
The stresses \sX, \sZ~and \sXZ~at the same point of the solid are represented by the continuous black, dotted gray and black dashed lines, respectively. The arrows point out the peaks created by the convergence of the different waves. Notice that \sXZ~is multiplied by a factor of 10, as the amplitudes are much smaller.

We observe that in the case of \sX~(continuous black line), at first there is a slight decrease to negative values due to the bulk wave convergence which induces (negative) stresses much smaller (in magnitude) than those of the Rayleigh wave. The inner ring bulk wave reaches \sX$=-0.16$\,GPa at $t=11$\,ns, and the one from the outer ring \sX$=-0.12$\,GPa at 17\,ns.

The stress reaches a significant positive peak (tension) when the Rayleigh waves converge, followed by a negative peak (compression). A similar trend is observed with \sZ, where the initial increase is moderate and the drop to negative values is similar to \sX. The case for the shear stress, \sXZ, has a similar trend to the pressure, which initially shows a small increase followed by a more significant decrease to negative values and an increase as the shock wave arrives to the center. 

In the case of the inner ring, the Rayleigh wave induces positive stress (tension) on the glass, with a maximum of \sX$=3.67$\,GPa at $t=19$\,ns. At $t=26$\,ns, as the shock wave reaches the center with a positive pressure peak of about 5\,GPa, it induces a negative peak of \sX$=-4.97$\,GPa. The corresponding minimum in \sZ$=-5.65$\,GPa is reached one nanosecond earlier, at $t=25$\,ns. 

The stresses induced by the outer ring reach a maximum of \sX$=2.14$\,GPa at $t=29$\,ns. The positive pressure peak of around 4\,GPa, at $t=42$\,ns induces negative peaks of \sX$=-3.86$\,GPa and \sZ$=-4.46$\,GPa. These stresses have the smallest amplitudes (considering all five studied cases), which could explain the large number of repetitions required to start observing damage ($a$ = 219 repetitions, mean value of the cumulative distribution function). 

\begin{figure*}
\includegraphics[width=\textwidth]{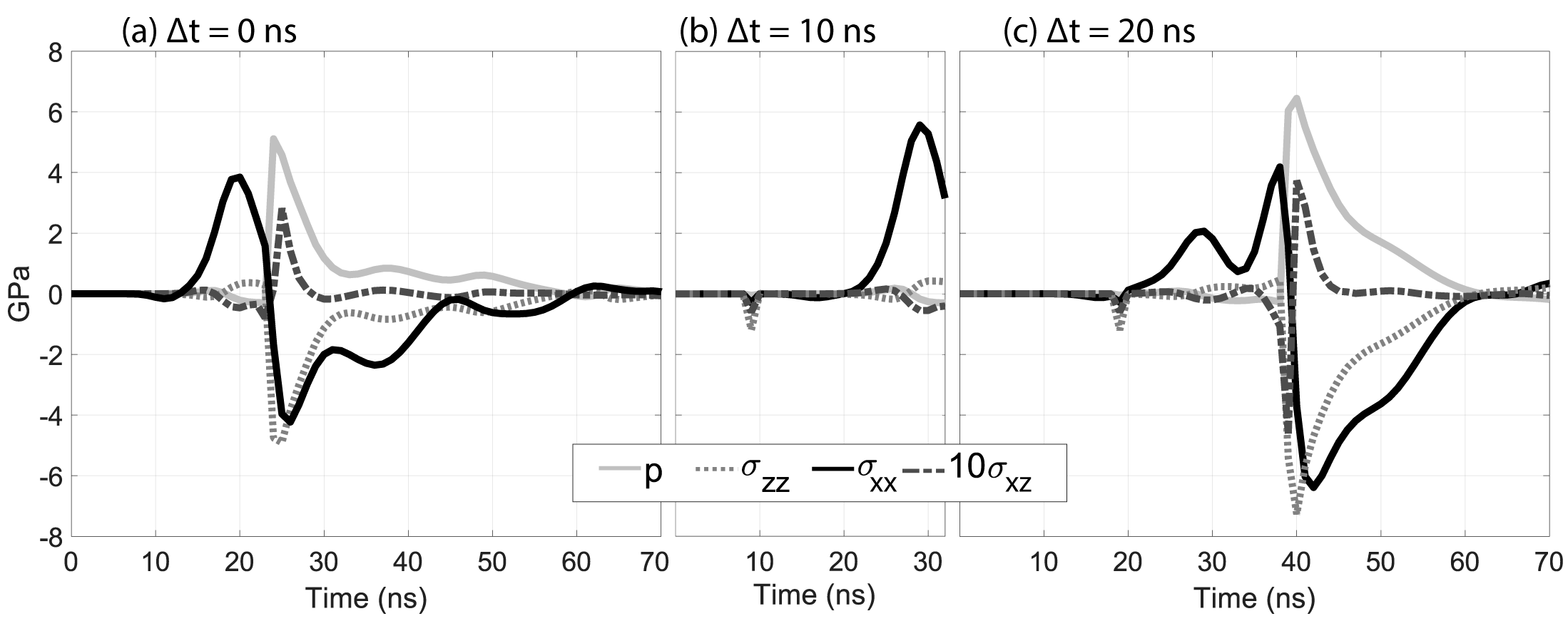}
  \caption{\label{fig: fig8}Calculated pressures and stresses at the center of the rings ($x=0$) at the solid/liquid interface ($z=0$). (a) \dt$=0$. (b) \dt$=10$\,ns. (c) \dt$=20$\,ns.}
\end{figure*}

\begin{table}
\begin{ruledtabular}
\begin{tabular}{ccccccc}
$\sigma$& &Minimum& & &Maximum&\\ 
  & value   & pos. & $t$ &  value   & pos.  & $t$ \\
  &  \scriptsize{(GPa)}  &  (\scriptsize{$\upmu$m})&  \scriptsize{(ns)}&   \scriptsize{(GPa)}  & \scriptsize{($\upmu$m)} & \scriptsize{(ns)}\\ \hline
 &\multicolumn{5}{c}{\bf{Inner}}\\\hline
 xx&-4.97&(0,0)& 26 & 3.67 &(0,0) & 19 \\
 zz&-5.65
 &(0,-3)&25&0.88
 &(0,-7)&20\\
 xz&-1.13&(11,-8)&20&1.31&(6,-6)&26
 \\ \hline
 &\multicolumn{5}{c}{\bf{Outer}}\\ \hline
xx&-3.86&(0,0)& 43 & 2.14 &(0,0) & 29 \\
 zz&-4.46
 &(0,-3)&43&0.61
 &(0,-10)&30\\
 xz&-0.60&(17,-19)&20&0.91&(5,-5)&44
 \\
\end{tabular}
\end{ruledtabular}
\caption{\label{tab:table2} Maximum and minimum stress values (\sX, \sZ, \sXZ), their position ($x$,$z$) and time of occurrence $t$ for the individual inner and outer ring. }
\end{table}


\subsection{Bullseye configuration}
Figure \ref{fig: fig8} shows the corresponding results for \dt$=0$, 10 and 20\,ns. The first column (a) contains the plot of pressures in the liquid and stresses at the solid surface at $x=z=0$ as a function of time. 

First we analyze the case of \dt$=0$\,ns (Fig. \ref{fig: fig8}a).
As in Fig. \ref{fig: fig7}, the pressure in the liquid is plotted with a gray continuous line. We observe a similar trend to that of the individual inner ring (Fig. \ref{fig: fig7}a).
The stresses \sX, \sZ~and \sXZ~are represented by a black continuous, gray dotted and black dashed line, respectively.

In the case of \sX~we observe a large increase when the inner Rayleigh waves arrive at the origin at $t=19$\,ns with a peak of 3.84\,GPa, followed by a rapid transition to a negative peak with a gradual relaxation and several smaller negative peaks. 
The first negative peak in \sX~is due to the arrival of the inner bulk wave, this is followed by a positive maximum when the inner Rayleigh wave converges, then a negative minimum is reached with the convergence of the inner shock wave, followed by other negative peaks due to the arrival of the outer Rayleigh and shock waves. 
\sZ~shows a smaller first positive peak than \sX, followed by a large drop to negative values, with small subsequent variations during relaxation.

We note that for \dt$=0$ the minimum has a value of -4.43\,GPa, while for the individual inner ring it reaches -4.98\,GPa. That might explain the less destructive effect of \dt$=0$ compared with the individual rings once the substrate is damaged (Fig. \ref{fig: fig4}).

The case of \dt$=10$\,ns shows the largest peak in \sX~among all cases (5.56\,GPa) as the Rayleigh waves from both sources converge in the center at the same time (Fig. \ref{fig: fig8}(b)), then the code seizes to converge shortly after the pressure becomes negative (with a larger amplitude than for the individual rings).
We suspect that this case should achieve the strongest negative stresses because with this delay time we observe the maximum amount of damage (Fig. \ref{fig: fig4}) in the experiments.
To avoid the numerical model collapse, we explored adjusting different parameters including finer meshes, smaller time steps and others. However, in all cases, the simulation does not converge for delay times \dt between 8 and 16\,ns. This might be explained by a sudden decrease of the stresses (if it follows the trends for the other delay times), preventing convergence of the elasticity equation.
The experimental results for \dt$=10$\,ns (Fig. \ref{fig: fig4}) show that only after 10 repetitions, there is significant glass damage. After 37 repetitions, the damage is extensive and reaches the opposite side of the glass cover slip (height of 160\mm). 

The case of \dt$=20$\,ns (Fig. \ref{fig: fig8}c) corresponds to simultaneous convergence of the shock waves, so the increase in pressure in the liquid appears at a later time as expected, with a larger amplitude.
\sX~has two main positive peaks: one at $t=29$\,ns due to the arrival of the outer Rayleigh wave and another at $t=38$\,ns induced by the inner Rayleigh wave. Then, as the pressure in the liquid increases to 6.44\,GPa, \sX~shows a rapid decrease from 4.38\,GPa at $t=38$\,ns to $-6.38$\,GPa at $t=42$\,ns, staying negative for about 20\,ns. The case of \sZ~again shows a moderate increase, followed by a decrease to a negative peak of $-7.26$\,GPa.
The plot or \sXZ~(dotted black line) shows an abrupt change from negative to positive values, from $-0.46$\,GPa at $t=39$\,ns to 0.37\,GPa at $t=40$\,ns. 

In another study \cite{Veysset2017}, using a gold coated glass substrate, the authors mentioned that it was possible that the observed glass delamination is caused by the rapid changes in $\sigma$, \sZ~achieves the largest negative values, but \sX~undergoes the largest changes in amplitude (positive/negative) as it reaches large positive values and swings to large negative values. Those results are consistent with our simulations and the experimental results.
Another important observation is that while the maximum/minimum values of \sX~are reached at $x=z=0$, the largest amplitudes for \sZ~are reached at $x=0$ inside the glass at heights $z$ between $-3$\mm~and $-10$\mm. In the case of \sXZ, the maximum/minimum values are reached inside the glass as well, at heights ranging from $-19$\mm~to $-5$\mm, and at horizontal positions $x$ between 17\mm~and 5\mm. 
The tables \ref{tab:table2} and \ref{tab:table3} list the positions and times at which the maximum and minimum stresses are reached for each case. 

Out of the delay times (8-16\,ns) for which the code does not converge, the maximum positive amplitude (5.81\,GPa) is reached for \dt$=13$\,ns.

\begin{table}
\begin{ruledtabular}
\begin{tabular}{ccccccc}
$\sigma$& &minimum& & &maximum&\\ 
  & value   & pos. & $t$ &  value   & pos.  & $t$ \\
  &  \scriptsize{(GPa)}  &  (\scriptsize{$\upmu$m})&  \scriptsize{(ns)}&   \scriptsize{(GPa)}  & \scriptsize{($\upmu$m)} & \scriptsize{(ns)}\\  \hline
 &\multicolumn{5}{c}{\bf{$\Delta$t = 0\,ns}}\\\hline
 xx&-4.22&(0,0)& 26 & 3.84 &(0,0) & 20\\
 zz&-5.90 &(0,-3)&25 &0.88 &(0,-9)&19\\
 xz&-1.10&(11,-7)&20&1.30&(5,-5)&26
 \\ \hline
 &\multicolumn{5}{c}{\bf{$\Delta$t = 10\,ns}}\\ \hline
 xx& & &   & 5.56 &(0,0) & 30 \\
 zz&  & & &1.30 &(0,-9)&29\\
 xz& &  &  &  &  &
 \\ \hline
 &\multicolumn{5}{c}{\bf{$\Delta$t = 20\,ns}}\\ \hline
 xx&-6.38&(0,0)& 42 & 4.18 &(0,0) & 38 \\
 zz&-8.46 &(0,-3)&40&0.81 &(0,-5)&37\\
 xz&-1.50&(6,-5)&38&1.92&(7,-8)&42
 \\
\end{tabular}
\end{ruledtabular}\caption{\label{tab:table3} Maximum and minimum stress (\sX, \sZ, \sXZ) values at a given position ($x$,$z$) and time for \dt$=0$, 10 and 20\,ns.
}
\end{table}

\section{Conclusion}

Two cylindrical waves launched at different radii from a common center when superpositioned constructively significantly increase the magnitude of stresses as compared to a single focused wave. We have termed this focusing strategy of two cylindrical waves “Bullseye focusing”.

In the experiments, the initial radius of the two cylindrical waves was kept constant, yet the delay between their launches was varied. Here, Bullseye focusing works best for a delay of \dt$=10$\,ns, where the largest damage to the substrate is obtained, also with the lowest number of runs. The simulations reveal for this delay the highest tension on the substrate due to the focused Rayleigh waves. Here the simulations did not converge, yet the highest compressive stresses were found for a delay of \dt= 20\,ns. Utilizing a more advanced model for the glass may allow to overcome the convergence problem. Then we expect that the highest compressive stresses in the simulations also for the delay of \dt$=10$\,ns.

Interestingly, for a delay of \dt$=0$ the same number of runs as for a single inner ring is necessary to cause damage, still the superposition two waves has an effect: the area of the central damage. There the damaged area is considerably smaller than for the case of a single inner ring. We explain this observation with a partial destructive interference of the two Rayleigh waves. 

In all simulated cases, we observe abrupt changes in the sign of the stress components \sX~and \sZ. The biggest positive peak is caused by the arrival of the Rayleigh waves at the center (tension) while the following negative peak that is caused by the arrival of the shock wave (compression). In the case of \sXZ, the trend is reversed; the first peak due to Rayleigh waves convergence compresses the substrate, while the second, generated by the convergence of shock wave, is applying a tension. These peaks last for a few nanoseconds and drop from positive to negative  within a few nanoseconds. The induced stresses range between $5.56\,$GPa and -$6.38\,$GPa. 

The concept of Bullseye focusing as presented here is not limited to two rings. It can be extended to more rings to potentially achieve even larger stresses at their common focus. For example, adding a third pulsed laser and illuminating a separate region of the spatially multiplexed digital hologram would give rise to a third converging wave with arbitrary timing. The finding that the size of the damage can be controlled, too, could be another advantage of the present focusing scheme. Here we envision using the setup to explore the possibility to control the damage size to the substrate with the idea to create a Bullseye cutter for glass.

The simulations of the fluid-structure interaction can be improved, too, e.g., through a more realistic model for the structural deformation that accounts for nonlinearities and plastic deformation and by implementing more accurate numerical schemes to improve convergence.

Overall we have demonstrated a comparatively simple method to generate stresses in solids that can easily overcome the yield strength of many materials. Enhancing the amplification of the stresses and utilization for cutting of brittle materials are two straightforward extensions of the Bullseye focusing technique.

\begin{acknowledgments}
Thanks to Jose Rangel for machining components and to Cristian Mojica-Casique for programming and arduino controlled flip mount.
This work is partially funded by DGAPA UNAM PAPIIT grant IN107222; CTIC-LANMAC; CONACYT LN-299057; and the DFG (German Research Association) under contract OH 75/4-1.
\end{acknowledgments}

\bibliography{main}

\end{document}